\documentstyle[12pt]{article}

\def\thefootnote{\fnsymbol{footnote}}
\def\re{{\rm Re}}
\def\im{{\rm Im}}
\def\s{\bar{s}}
\def\t{\bar{t}}
\def\tG{{\tilde G}} 

\def\[{\left [}
\def\]{\right ]}
\def\({\left (}
\def\){\right )}
\def\bl{\bar{\lambda}}

\def\pp{\partial}

\def\G{{\cal G}}

\newcommand{\prl}{{\it Phys.\ Rev.\ Lett. }}
\newcommand{\pr}{{\it Phys.\ Rev. }}
\newcommand{\pl}{{\it Phys.\ Lett. }}
\newcommand{\np}{{\it Nucl.\ Phys. }}
\newcommand{\be}{\begin{equation}}
\newcommand{\ee}{\end{equation}}
\newcommand{\bea}{\begin{eqnarray}}
\newcommand{\eea}{\end{eqnarray}}

\newcommand{\WaWa}{W_a^{\alpha}W^a_{\alpha}}

\newcommand{\DbDb}{{\cal D}_{\dot{\alpha}}{\cal D}^{\dot{\alpha}}}

\catcode`\@=11

\begin{document}
\begin{titlepage}
\begin{center}
      \hfill  LBNL-41802 \\
      \hfill  UCB-PTH-98/25 \\
      \hfill hep-ph/9805320\\
\hfill May 1998
\vskip .5in

{\large \bf Phenomenology and cosmology of weakly coupled string 
theory }\footnote{Talk presented at The Richard Arnowitt Fest, April 5-8, 1998,
Texas A \& M University, College Station, TX, to be published in the 
proceedings.}\footnote{This work was 
supported in part by the Director, Office of
Energy Research, Office of High Energy and Nuclear Physics, Division
of High Energy Physics of the U.S. Department of Energy under
Contract DE-AC03-76SF00098 and in part by the National Science
Foundation under grant PHY-95-14797.}\\[.1in]

Mary K. Gaillard

{\em Department of Physics,University of California, and

 Theoretical Physics Group, 50A-5101, Lawrence Berkeley National Laboratory,
   Berkeley, CA 94720, USA}
\end{center}

\begin{abstract}
The weakly coupled vacuum of $E_8\otimes E_8$
heterotic string theory remains an attractive
scenario for phenomenolgy and cosmology.  The particle spectrum is 
reviewed and the issues of gauge coupling unification, dilaton stabilization 
and modular cosmology are discussed.  A specific model for condensation and
supersymmetry breaking, that respects known 
constraints from string theory and is phenomenologically viable, is described.

\end{abstract}
\end{titlepage}
\renewcommand{\thepage}{\roman{page}}
\setcounter{page}{2}
\mbox{ }

\vskip 1in

\begin{center}
{\bf Disclaimer}
\end{center}

\vskip .2in

\begin{scriptsize}
\begin{quotation}
This document was prepared as an account of work sponsored by the United
States Government. Neither the United States Government nor any agency
thereof, nor The Regents of the University of California, nor any of their
employees, makes any warranty, express or implied, or assumes any legal
liability or responsibility for the accuracy, completeness, or usefulness
of any information, apparatus, product, or process disclosed, or represents
that its use would not infringe privately owned rights. Reference herein
to any specific commercial products process, or service by its trade name,
trademark, manufacturer, or otherwise, does not necessarily constitute or
imply its endorsement, recommendation, or favoring by the United States
Government or any agency thereof, or The Regents of the University of
California. The views and opinions of authors expressed herein do not
necessarily state or reflect those of the United States Government or any
agency thereof of The Regents of the University of California and shall
not be used for advertising or product endorsement purposes.
\end{quotation}
\end{scriptsize}

\vskip 2in

\begin{center}
\begin{small}
{\it Lawrence Berkeley Laboratory is an equal opportunity employer.}
\end{small}
\end{center}

\newpage
\renewcommand{\thepage}{\arabic{page}}
\setcounter{page}{1}
\def\thefootnote{\arabic{footnote}}
\setcounter{footnote}{0}

\section{Introduction}

It has been suggested that weakly coupled string theory has serious
difficulties~\cite{diff,bd1}. 
Specifically, arguments have been made that {\it i}) dilaton 
stabilization is not possible, {\it ii}) the prediction for coupling constant 
unification is incompatible with experiment, and {\it iii}) there are serious
cosmological problems.  In this talk I address these issues, and show
that in the context of a specific model, consistent with the known constraints
from string theory, a weakly coupled vacuum presents a viable scenario.

First recall the reasons for the original appeal of the weakly coupled
$E_8\otimes E_8$ heterotic string theory~\cite{gross} compactified on a 
Calabi-Yau (CY) manifold~\cite{cy} (or a CY-like orbifold~\cite{orb}).  The
zero-slope (infinite string tension) limit of superstring theory~\cite{gs} 
is ten dimensional supergravity coupled to a supersymmetric Yang-Mills theory
with an $E_8\otimes E_8$ gauge group.  To make contact with the real world,
six of these ten dimensions must be compact -- of size much smaller than
distance scales probed by particle accelerators, and generally assumed to be of
order of the reduced Planck length, $10^{-32}$cm.  If the topology of
the extra dimensions were a six-torus, which has a flat geometry, the
8-component spinorial parameter of $N=1$ supergravity in ten dimensions would
appear as the four two-component parameters of $N=4$ supergravity in ten
dimensions.  On the other hand, a Calabi-Yau manifold leaves only one of
these spinors invariant under parallel transport; for this manifold the group
of transformations under parallel transport (holonomy group)
is the $SU(3)$ subgroup of the
maximal $SU(4) \cong SO(6)$ holonomy group of a six dimensional compact space. 
This breaks $N=4$ supersymmetry to $N=1$ in four dimensions.  
As is well known, the only phenomenologically viable
supersymmetric theory at low energies is $N=1$, because it is the only one that
admits complex representations of the gauge group that are needed to describe
quarks and leptons. For this solution,
the classical equations of motion impose the identification of the affine
connection of general coordinate transformations on the compact space 
(described by three complex dimensions) with the gauge connection of an 
$SU(3)$ subgroup of one of
the $E_8$'s: $E_8\ni E_6\otimes SU(3)$.  As a consequence the gauge group in
four dimensions is $E_6\otimes E_8$. Since the early 1980's, $E_6$ has been 
considered the largest group that is a phenomenologically viable candidate for 
a Grand Unified Theory (GUT)
of the Standard Model (SM).  Hence $E_6$ is identified as the 
gauge group of the ``observable sector'', and the additional $E_8$ is attributed
to a ``hidden sector'', that interacts with the former only with gravitational
strength couplings.

Orbifolds, which are flat spaces except for points of infinite curvature, are
more easily studied than CY manifolds, and orbifold compactifications that
closely mimic the CY compactification described above, and that yield
realistic spectra with just three generations of quarks and leptons, have been
found~\cite{iban}.  In this case the surviving gauge group is $E_6\otimes G_o
\otimes E_8,\;G_o\in SU(3)$. 

The low energy effective field theory is determined by the 
massless spectrum, that is the spectrum of states with masses very small
compared with the scales of the string tension and of compactification. Massless
bosons have zero triality under an $SU(3)$ which is the diagonal of the
$SU(3)$ holonomy group and the (broken) $SU(3)$ subgroup of one $E_8$.
The ten-dimensional vector fields $A_M,\; M = 0,1,\ldots 9,$ appear in 
four dimensions as four-vectors $A_\mu,\;\mu = M = 0,1,\ldots 3$, and as 
scalars $A_m,\; m = M-3 = 1,\cdots 6.$  Under the decomposition
$E_8\ni E_6\otimes SU(3)$, the $E_8$ adjoint contains the adjoints of $E_6$ and
$SU(3)$, and the representation 
${\bf(27,3)} + {\bf(\overline{27},\overline{3})}$.  Thus the massless 
spectrum includes gauge fields in the adjoint representation of 
$E_6\otimes G_o\otimes E_8$ with zero triality under both $SU(3)$'s, 
and scalar fields in ${\bf 27 + \overline{27}}$ of $E_6$, with triality $\pm1$ 
under both $SU(3)$'s, together with their fermionic superpartners.  
The number of ${\bf 27}$'s and ${\bf\overline{27}}$ chiral supermultiplets 
that are massless depends on the detailed topology of the compact manifold.  
The important point for phenomenology is the 
decomposition under $E_6\to SO(10)\to SU(5)$:
\be ({\bf 27})_{E_6} = ({\bf 16 + 10 + 1})_{SO(10)} = 
\({\bf \{\bar{5} + 10 + 1\} + \{5 + \bar{5}\} + 1}\)_{SU(5)}.\ee
A ${\bf \overline{5} + 10 + 1}$ contains one generation of 
quarks and leptons of the 
Standard Model, a right-handed neutrino and their scalar superpartners; a 
${\bf 5 + \overline{5}}$ contains the two Higgs doublets needed in the
supersymmetric extension of the Standard Model and their fermion superpartners, 
as well as color-triplet supermultiplets. Thus all the states
of the Standard Model (and its minimal supersymmetric extension) are present.

On the other hand, there are no scalar particles in the adjoint
representation of the gauge group. In conventional models for grand unification,
these (or one or more other representations much larger than the fundamental 
one) are needed to break the GUT group to the Standard Model.  In string 
theory, this symmetry breaking can be achieved by the Hosotani, or ``Wilson 
line'', mechanism~\cite{hos} in which gauge flux is trapped around ``holes'' or
``tubes'' in the compact manifold, in a manner reminiscent of the Arahonov-Bohm 
effect.  The vacuum value of the trapped flux $<\int d\ell^m A_m>$ has the same
effect as an adjoint Higgs, without the complications of having to construct a
potential for large Higgs representations that can actually reproduce the
properties of the observed vacuum~\cite{lang}.
When this effect is included, the gauge group in four dimensions is 
\bea &&\G_{obs}\otimes\G_{hid}, \quad\G_{obs}=\G_{SM}\otimes\G'\otimes\G_o,\quad
\G_{SM}\otimes\G'\in E_6, \quad \G_o\in SU(3),\nonumber \\ &&
\G_{hid}\in E_8,\quad \G_{SM} = SU(3)_c\otimes SU(2)_L\otimes U(1)_w.
\label{group}\eea

There are of course many other four dimensional string vacua besides the class
of vacua described above.

\noindent $\bullet$ The gauge group in four dimensions may be larger. 
Its rank $r$ can be greater than the rank 16 of $E_8\otimes E_8$, if some
Kaluza-Klein vector fields $g_{\mu m}$ are massless. In weakly coupled string 
theory $r\le 22$, but the group can become arbitrarily large in the strongly 
coupled regime~\cite{cand}.  These scenarios, however, seem to lead further 
away from observation.

\noindent $\bullet$ The above scenario corresponds to affine level $k_a=1$,
where $a$ refers to the gauge group $\G_a$, and the affine levels $k_a$ are the 
coefficients of Schwinger terms in the current algebra on the string world 
sheet.  For nonabelian groups $k_a$ is a positive integer.
If $k_a>1$, it is possible to have adjoint Higgs multiplets in the low 
energy theory, offering the possiblity of a conventional GUT below the 
Planck scale (with the problems alluded to above). Models with $k_a\ne 1$ have 
proven difficult to construct, although some examples exist.

The attractiveness of the picture described above is that the requirement of
$N=1$ supersymmetry (SUSY) naturally results in a phenomenologically viable
gauge group and particle spectrum. Moreover, the gauge symmetry can be broken 
to a product group embedding the Standard
Model without the necessity of introducing large Higgs representations.

Supersymmetry is broken in nature. It is well known that spontaneous breaking
of global supersymmetry in the observable sector is incompatible with the
observed low energy mass spectrum. This fact led Arnowitt and Nath, among
others, to the formulation~\cite{sugra} of spontaneously broken supergavity by 
``hidden sector'' interactions that communicate with the observable sector
{\it via} gravitational strength couplings that induce soft SUSY breaking
terms in the effective low energy theory, assumed to be a 
supersymmetric extension of the Standard Model.  The $E_8\otimes E_8$ string 
theory provides us with the needed hidden sector.

More specifically, if some subgroup $\G_a$ of $\G_{hid}$ is asymptotically free,
with a $\beta$-function coefficient $b_a>b_{SU(3)}$, defined by the 
renormalization group equation (RGE) 
\be \mu{\pp g_a(\mu)\over\pp\mu} = -{3\over2}b_ag_a^3(\mu) + 
O(g_a^5)\label{rge},\ee
confinement and fermion condensation will occur at a scale
$\Lambda_c\gg\Lambda_{QCD}$, and hidden sector gaugino condensation 
$<\bl\lambda>_{\G_a} \ne 0,$ may induce~\cite{nilles} supersymmetry breaking.

To discuss supersymmetry breaking in more detail, 
we need the low energy spectrum resulting from the ten-dimensional gravity
supermultiplet that consists of the 10-d
metric $g_{MN}$, an antisymmetric tensor $b_{MN}$, the dilaton $\phi$, the
gravitino $\psi_M$ and the dilatino $\chi$.  For the class of CY and orbifold
compactifications described above, the massless bosons in four dimensions are
the 4-d metric $g_{\mu\nu}$, the antisymmetric tensor $b_{\mu\nu}$, the dilaton
$\phi$, and certain components of the tensors $g_{mn}$ and $b_{mn}$ that form
the real and imaginary parts, respectively, of complex scalars known as moduli;
the number of moduli is related to the number of particle generations (\# of
${\bf 27}$'s $-$ \# of ${\bf\overline{27}}$'s).  (More precisely, the scalar
components of the chiral multiplets of the low energy theory are obtained as 
functions of the scalars $\phi,g_{mn}$ while the pseudoscalars $b_{mn}$ form
axionic components of these supermultiplets.) Typically, in a three 
generation orbifold model there are three moduli $t_I$; the $vev$'s 
$<{\rm Re}t_I>$ determine the radii
of compactification of the three tori of the compact space.  In some
compactifications there are three other moduli $u_I$; the $vev$'s 
$<{\rm Re}u_I>$ determine the ratios of the two {\it a priori} independent 
radii of each torus.  These form chiral multiplets with fermions $\chi^t_I,
\chi^u_I$ obtained from components of $\psi_m$.
The 4-d dilatino $\chi$ forms a chiral multiplet with with a complex scalar 
field $s$ whose $vev$
\be <s> = g^{-2} - {i\over8\pi^2}\theta \label{dil}\ee
determines the gauge coupling constant and the $\theta$ parameter of the 4-d
Yang-Mills theory.  The ``universal'' axion Im$s$ is obtained by a duality
transformation~\cite{wit} from the antisymmetric tensor $b_{\mu\nu}: \;
\pp_\mu{\rm Im}s\leftrightarrow
\epsilon_{\mu\nu\rho\sigma}\pp^\nu b^{\rho\sigma}.$

Because the dilaton couples to the (observable and hidden) Yang-Mills
sector, gaugino condensation induces~\cite{dine} a superpotential for the
dilaton superfield\footnote{Throughout I use capital Greek or Roman letters
to denote a chiral superfield, and the corresponding lower case letter to denote
its scalar component.} $S$: 
\be W(S) \propto e^{-S/b_a}.\ee
The vacuum value
\be <W(S)> \propto \left<e^{-S/b_a}\right> = e^{-g^{-2}/b_a}= \Lambda_c,\ee
is governed by the condensation scale $\Lambda_c$ as determined by the
RGE (\ref{rge}).  If it is nonzero, the gravitino 
$\tG$ acquires a mass $m_{\tG}\propto<W>$, and local supersymmetry is broken. 

\section{Gauge coupling unification}
Precision data on the three gauge couplings of the Standard Model is often
construed as indirect evidence for supersymmetry.  Using the measured values
of these couplings at the $Z$ mass, the RGE equations applied to
the SM give approximate, but not exact unification at a
scale around $10^{15}$GeV, whereas in the minimal supersymmetric extension
of the SM (MSSM), the data are consistent with exact unification at the
scale $\Lambda_G \approx 2\times 10^{16}$GeV, with a value of the fine structure
constant $\alpha_G = g^2_G/4\pi\approx 1/25$.  As discussed above,
string theory is not necessarily a GUT, but all the coupling constants
are determined -- at a scale $\mu_0$ characteristic of the underlying string 
theory -- by the $vev$ (\ref{dil}).  Allowing for affine levels $k_a\ne 1$, the 
prediction is
\be g^{-2}_a(\mu_0) = k_a<{\rm Re}s>. \label{unif}\ee
There are several possibilities within the general context of weakly coupled
superstring theory:

\noindent $\bullet$ $k_a = 1\;\forall \;a$, in which case SM unification is
predicted\footnote{I use the GUT normalization for $U(1)$, not the SM
normalization which is sometimes used in the literature.}  
as in conventional GUTs, but the theory above the unification scale is not
a GUT field theory.

\noindent $\bullet$ $k_a \ne 1$ in which case there are two distinct scenarios.

{\it i})  Since Higgs superfields in the adjoint representation
can appear in the effective field theory if $k_a>1$, this theory may be a 
GUT that is broken to the SM by a Higgs $vev$ which is determined by
the dynamics of the field theory.  One recovers the conventional SUSY GUT
scenario (with its conventional difficulties) and string theory provides
no additional constraint.

{\it ii}) The theory may not be a GUT, and the RGE prediction is modified
if the $k_a$ of the three SM gauge groups are not the same.

What distinguishes string theory from conventional GUTs is that the scale 
$\mu_0$ of unification is not an arbitrary parameter, but is determined in terms
of the Planck scale by one or more scalar $vev$'s, since the theory contains
only one fundamental scale: the string tension $m_s^{-2}$, related to the 
reduced Planck mass $m_P= (8\pi G_N)^{-{1\over2}} \approx 2\times 10^{18}$GeV 
by $m^2_s = m^2_P/<{\rm Re}s> = g^2m_P^2$.  An educated guess~\cite{wit} would 
be to identify $\mu_0$ with the scale of compactification:
\be \Lambda_{comp} = \left<({\rm Re}t)^{-{1\over2}}\right>m_s =
\left<({\rm Re}t{\rm Re}s)^{-{1\over2}}\right>m_P,\label{comp}\ee
where $t$ is the geometric mean of the moduli $t_I$.
Then comparison with the data assuming affine level one would yield
$<{\rm Re}s>\approx2,\;<{\rm Re}t>\approx 50$. It has been argued by
Kaplunovski~\cite{kap0} that such a large value of $<t>$ is not possible; 
he concluded that consistency requires $t\sim 1,\;\mu_0\sim
m_s$.  This argument was revisited~\cite{kap2} in the light of recent progress
in strongly coupled string theory (M-theory~\cite{duff}).  The conclusion was
that large $t$ (small radius of compactification) is excluded in most string
vacua, a notable exception being a particular strongly coupled limit~\cite{hw}
of M-theory in which there is an eleventh dimension much larger than the
ten-dimensions on which string theory lives.

However, it is possible to be more precise, particularly in the case of
well-understood orbifold compactifications.  The field theory loop corrections
must be calculated using a regularization procedure that respects supersymmetry.
Using a Pauli-Villars 
(PV) regularization\footnote{The cancellations present in
supersymmetric theories allow one to regulate gauge loops with Pauli-Villars
fields in chiral supermultiplets~\cite{mk}, and BRST invariance is maintained.}
one obtains for the loop-corrected gauge couplings~\cite{tom} (in the sense of 
the ``Wilson coefficient'' of the Yang-Mills field strength operator 
$-{1\over4}F^a_{\mu\nu}F_a^{\mu\nu})$:
\bea g_a^{-2} &=& k_a{\rm Re}s + \sum_I{\ln\(t_I + \t_I\)\over16\pi^2}
\[C^a - \sum_\alpha(1-2q^I_\alpha)C^a_\alpha\] + {\ln(s + \s)\over16\pi^2}
\(C^a - \sum_\alpha C^a_\alpha\)\nonumber \\ &\equiv& 
k_a{\rm Re}s - 3C^a{\ln\Lambda_a^2\over16\pi^2} +
\sum_\alpha C^a_\alpha{\ln\Lambda_\alpha^2\over16\pi^2}, \quad
b_a = {1\over8\pi^2}\(C^a - {1\over3}\sum_\alpha C^a_\alpha\),\label{loop}\eea
where $q^I_\alpha$ is a ``modular weight'', 
$C^a$ is the quadratic Casimir operator in the adjoint representation
of the gauge group $\G_a$ and $C^a_\alpha = {\rm Tr}(T^a_\alpha)^2$, where
$T^a_\alpha$ represents a generator of $\G_a$ on the 
matter chiral superfield $\Phi^\alpha$.
On the RHS the $\Lambda$'s are the PV masses that act as effective cut-offs 
(which determine the reference value $\mu_0$ in the RGE equations) for the 
different sectors of the theory in a one-loop calculation.  For the matter 
fields of
the ``untwisted sector'' in orbifold compactification, $\alpha = (AI), \;
q^I_\alpha = q^I_{AJ} = \delta^I_J$, the effective cut-off turns out to be
precisely (\ref{comp}) if $<t_I> = <t>$.  For the gauge sector, the effective 
cut-off is $\Lambda_a = g^{-{2\over3}}\Lambda_{comp}$; the factor 
$g^{-{2\over3}} = <(\re s)^{1\over3}>$ corresponds to a two-loop correction that
appears automatically in a one-loop calculation if the PV masses are chosen so
as to respect supersymmetry.  For the remaining (``twisted sector'')
contributions, different powers of the moduli appear in the cut-off, but the
anticipated result that the effective cut-off for the low energy field theory is
determined by the moduli is indeed borne out.

However, (\ref{loop}) cannot be the correct answer.  The result is not invariant
under the group of modular transformations generated by 
\be T_I\to T_I^{-1}, \quad T_I\to T_I + i \label{mod} \ee
(together with $q^I_\alpha$-dependent transformations on $\Phi^\alpha$), 
that is known~\cite{mod} to be an exact symmetry of string
perturbation theory.  The effective field theory has a conformal anomaly (due 
to the noninvariance of the cut-off) and a chiral anomaly (due to the 
chiral transformation of fermion fields implicit in the superfield
transformations) under (\ref{mod}).  
These anomalies form a supermultiplet, a constraint that fixes~\cite{tom} the
cut-offs $\Lambda$ in (\ref{loop}), since the chiral anomaly is unambiguously 
determined at one loop in quantum field perturbation theory.

The upshot is that one must add counterterms to the effective field theory to
restore modular invariance.  In general two different counterterms provide the
anomaly cancellation. Some or all of the variation under (\ref{mod}) of the 
field theory loop contributions to $g^{-2}(\mu_0)$ can be canceled if $S$ is 
not modular invariant, resulting in a compensating variation of the tree level 
contribution [the first term on the RHS of (\ref{loop})].
This is the so-called Green-Schwarz counterterm~\cite{gsterm} which is model
independent; in the chiral formulation for the dilaton supermultiplet it has 
the effect of modifying the K\"ahler potential for the dilaton by a 
$T_I$-dependent term so as to maintain modular invariance of the K\"ahler 
potential.  In addition there are model-dependent
threshold corrections~\cite{thresh} that arise from integrating out the heavy
string and Kaluza-Klein modes; these generate terms in the loop-corrected value
of $g^{-2}(\mu_0)$ involving the modular invariant function 
$|\eta(t_I)|^4\re t_I$, where $\eta(t_I)$ is the Dedekind eta-function. 
Matching~\cite{tom} field theory loop calculations to string loop
calculations gives the boundary condition for the running coupling constants 
in the $\overline{MS}$ renormalization scheme:\footnote{Not included here is 
an additional moduli-dependent contribution from $N=2$ sectors that is
independent of the gauge group~\cite{kkpr}.}
\bea g_a^{-2}(\mu_0) &=& {k_a\over2\ell} - {C_a\over8\pi^2}\ln2 - 
{1\over16\pi^2}\sum_Ib^I_a\ln\(2\re t_I|\eta(t_I)|^4\), \nonumber \\
\mu_0^2 &=& <e^{-1}\ell> = {g^2\over2 e} = {m_s^2\over2 e}\label{gstring}\eea
in reduced Plank units $m_P=1$, where 
$\ell$ is a modular invariant function of the dilaton and the moduli:
\be \ell = \[2\re s - {C_{E_8}\over8\pi^2}\sum_I\ln(2\re t_I)\]^{-1}.\ee 
The same result was obtained in~\cite{kl} using a different regularization
procedure for field theory loops (and hence a different definition of the
Wilson coefficient).  The parameters
\be b_a^I = C_{E_8} - \sum_\alpha\(1 - 2q_I^\alpha\)C_a^\alpha  \ee
vanish~\cite{thresh} for a large class of orbifolds, in particular the 
$Z_3,Z_7$ orbifolds that appear to yield realistic models~\cite{iban}.  

The scalar field $\ell$ and the two-form $b_{\mu\nu}$ described in
the introduction are the bosonic 
components of a linear supermultiplet~\cite{linear} $L$, that is
dual to the chiral multiplet $S$.  The introduction of the Green-Schwarz 
counterterm is most naturally implemented within this formalism~\cite{bggm}.
There is increasing evidence~\cite{deboer} that this is the appropriate
formulation for the string dilaton.  This formulation is used in the explicit
model~\cite{us} for gaugino condensation and dilaton stabilization to be
discussed in the next section.  Although it has been argued~\cite{bdqq} that 
the two formalisms (linear and chiral) are equivalent even in the presence of
nonperturbative effects like gaugino condensation, the condensate action is 
more simply expressed in the linear multiplet formalism.

The bottom line of the present analysis is that for affine level one models
without significant threshold corrections [the last term in the RHS of
(\ref{gstring})], the unification scale $\mu_0$ is related to the value
of the common coupling constant at that scale by $\mu_0 \approx g M_P$,
as originally found by Kaplunovski~\cite{kap}.  
While the result (\ref{gstring}) has been derived only for orbifold
compactifications, its large $t_I$ limit agrees with the behavior found 
in the large $t_I$ limit of Calabi-Yau compactification. If one compares
the MSSM fit to the data of the unification scale 
($\mu_0 \approx 2\times 10^{16}$GeV) with the value obtained using the string
prediction and the fit value of the coupling at unification [$\alpha \approx
1/25,\; g^2\approx 0.5\Rightarrow \mu_0 \approx (2e)^{-{1\over2}}gm_P \approx
6\times 10^{17}$GeV], there is a mismatch of a factor of about 30 in $\mu_0$
[or a factor of about 3.4 in $\ln(\mu_0)$, which is the quantity that enters
in the RGE].

The resolution of this discrepancy has been addressed by a number of
authors~\cite{unif}.  The options studied include

\noindent $\bullet$ String threshold corrections, {\it i.e.,} the last term in
(\ref{gstring}).  As noted above, these are absent in many orbifolds.  In
addition they are small if $<t_I>\sim 1$ as expected, and in most orbifolds 
they have the wrong sign to correct the discrepancy.

\noindent $\bullet$ Affine levels $k_a\ne 1$.  The data require that the ratio
of the $SU(3)_c$ and $SU(2)_L$ affine levels be close to unity.  Since these 
are integers, they must be very large integers.  The only known
models have $k_a = 1,2,3,$ (which is insufficient) and the dimensions of 
matter representations grows with $k_a$ -- taking us farther afield from the
observed spectrum. Models with $k_2 = k_3 = 1$ can be made to fit the data
provided that $k_1< 1$, which has few~\cite{aff} realizations in
actual orbifold compactifications.

\noindent $\bullet$ Non-MSSM chiral matter.  All known orbifold 
compactifications have additional chiral matter with respect to the MSSM. 
These transform according to real (reducible) representations of the SM gauge 
group: $\({\bf r + \bar{r}}\)_{SM}\in \({\bf 27 + \overline{27}}\)_{E_6}$, and 
thus
can acquire gauge invariant masses well above the scale of electroweak symmetry
breaking.  By including such states with masses below the string string scale,
(\ref{gstring}) can be brought into agreement with the data with weak
coupling at the string scale $g(m_s) = O(1)$.  Fits to the data require
one or more additional states transforming as ${\bf 3 + \overline{3}}$ under
$SU(3)_c$ as well as pairs of $SU(2)_L$ doublets, and/or 
states transforming as ${\bf (3,2) + (\overline{3},2)}$ under
$SU(3)_c\otimes SU(2)_L$; these fits can discriminate among models.  This 
appears to be the most straightforward and natural source of threshold 
corrections to an MSSM fit to the prediction (\ref{gstring}).

\section{Gaugino condensation and the runaway dilaton}

The superpotential (5) results in a potential for the dilaton of the form
\be V(s)\propto e^{-2\re s/b_a},\ee
which has its minimum at vanishing vacuum energy with $<\re s> \to\infty,\;
g^2\to 0$.  This is the runaway dilaton problem.  One possible way out is
the introduction of a second source of SUSY breaking such that the vacuum
energy vanishes, but the superpotential does not: $<W> = <W(S) + W'>\ne 0$.
Then the gravitino acquires a mass, and local SUSY is broken.  The only 
scenario of this type that has been realized explicitly is similar to the 
Hosotani mechanism: $W'$ is a constant induced~\cite{dine} 
by a $vev$ of the form
$<\int dv^{lmn}d_{[l}b_{mn]}>$, where $dv^{lmn}$ is a volume element
on the compact manifold, and $d_{[l}b_{mn]}$ is the curl of the anti-symmetric
tensor of 10-d supergravity.  The difficulty is that this $vev$ satisfies 
a quantization condition~\cite{rw}, which means that if $<W>\ne0$, the gravitino
mass will be near the Planck scale and a large hierarchy for local SUSY 
breaking cannot be generated (although the generation of a large hierarchy
for observable SUSY breaking is not {\it a priori} excluded~\cite{bg}).

When the Green-Schwarz term is included, a second dilaton runaway direction is
encountered.  The potential is no longer positive definite.  The small 
coupling ($\ell\to0$) behavior is unaffected (with $2\re s$ replaced by 
$\ell^{-1}$), but the potential has a maximum at $b_a\ell=  .5$,
and is negative for $b_a\ell> .5(1 + \sqrt{3})\approx 1.37$.  Since $b_a\le 
b_{E_8}\approx .38,$ $V$ is negative for $\alpha = \ell/2\pi > 1.32/b_a\pi > 
.57$. This is the strong coupling regime, and nonperturbative string effects 
cannot be neglected; they are expected~\cite{shenk} to modify the
K\"ahler potential for the dilaton, which in the perturbative limit
is $k(\ell)= \ln\ell$. It has been shown~\cite{us,casas} that these 
contributions can indeed stabilize the dilaton.

An explicit model based on affine level one\footnote{This is a simplifying but 
not a necessary assumption.} orbifolds with three untwisted moduli
$T_I$ and a gauge group of the form (\ref{group}) has been constructed.
The dilaton is taken to be the $\theta =\bar{\theta}=0$ component of a real
superfield $L$ that satisfies a modified linearity condition; {\it i.e.},
its chiral projection is
\be \(\DbDb - 8R\)L = \sum_a \WaWa + \sum_a U_a.\label{lin}\ee
The right hand side is a chiral superfield of chiral weight 2. The first term
is a sum over operators bilinear in the chiral superfields
$W^a_\alpha$ of the unconfined Yang-Mills sectors, while the second sum 
is over condensate superfields of the confined Yang-Mills sectors ({\it i.e.}, 
strongly coupled at scales $\Lambda_c^a$).  With this construction, the
condensate superfields 
automatically satisfy a constraint~\cite{bdqq,bgt,bg96} implied by the Bianchi 
identity, that is usually ignored in chiral supermultiplet formulations
of gaugino condensation.  The condensate self-couplings and their couplings to
confined matter consist of the classical contribution obtained by the
substitution $\WaWa \to u_a$ in the standard
Yang-Mills Lagrangian, a quantum field theory
correction obtained by anomaly matching~\cite{vy}, 
the Green-Schwarz term~\cite{bg96}
and string threshold corrections needed to restore modular invariance.

In this formalism it is convenient to introduce a function $f(\ell)$ that
modifies the string scale coupling constant\footnote{If one performs a duality 
transformation {\it via} a Lagrange multiplier~\cite{bggm}  
$S + \bar{S}$, the equations of motion for $L$ give $S + \bar{S} = 
[f(L)+1]/L$, and $g^{-2} = <\re s>$ in the chiral formulation of the classical
effective field theory with no GS term.} $g$, and is
related to $g(\ell) = k(\ell) - \ln\ell$ by a differential equation: 
\be g^2 = \left<{2\ell\over1 + f(\ell)}\right> , \quad
\ell g'(\ell)= -\ell f'(\ell) + f(\ell). \label{einstein}\ee 
The results of~\cite{shenk} suggest a parameterization of the form
\be f(\ell) = \sum_{n=0} a_n\ell^{-n/2}e^{-c_n/\sqrt{\ell}}.\label{param}\ee
Retaining just the first one or two terms in the expansion (\ref{param}),
the potential can be made positive definite everywhere and the parameters
can be chosen to fit two data points: the coupling constant $g^2\sim 1$ 
and the cosmological constant $\Lambda = 0$ (or very nearly so).  This is fine
tuning, but reasonable values can be obtained for the parameters, {\it e.g.},
$c_0 = c_1 = 1,\; a_1/a_0 <0$ with $a_0,a_1$ in the range 2--5 (positivity of
the potential requires $a_0>2$).

It should be emphasized that only one condensate $u_a$ is needed for
dilaton stabilization. This picture is very 
different from previously studied ``racetrack'' models~\cite{race} where 
dilaton stabilization is achieved through cancellations among different 
condensates with similar $\beta$-functions.  If more than one condensate
is present, nonperturbative corrections to the dilaton K\"ahler
potential are still required to stabilize the dilaton.  

If the gauge group for the dominant condensate (largest $b_a$)
is not $E_8$, the moduli $t_I$ are also stabilized through their couplings to
twisted sector matter and/or moduli-dependent string threshold corrections.
Their vacuum values~\cite{lyth} are at one of the two self dual 
points\footnote{The coefficients $c_n$ in (\ref{param})
were assumed to be moduli-independent.  If nonperturbative string  
contributions turn out to be moduli-dependent, and hence not modular invariant,
as found~\cite{eva} in a different orbifold from the class considered here,
the moduli $vevs$ would be slightly shifted from their self-dual points.}
in the fundamental domain(see, {\it e.g.}~\cite{iban}): $t^I = 1, e^{i\pi/6}$; 
hence the 4-d string theory is weakly coupled ($\re t\sim 1$), as well as the 
4-d field theory ($g^2\sim 1$). The moduli auxiliary fields vanish in the 
vacuum:
$<F^I> = 0$, avoiding a potentially dangerous source of flavor changing neutral
currents.  The nonholomorphic constraint on the condensate superfield $U_a$ 
implied by the chiral projection (\ref{lin}) is an essential ingredient in this 
last result~\cite{us}.

\section{Modular cosmology}

The soft SUSY breaking parameters were calculated in~\cite{us} for
$<t_I> = 1$; the results are similar if $<t_I> = e^{i\pi/6}$.
If there is only one condensate, the universal axion\footnote{In~\cite{us}
the condensate superfields $U_a$ are introduced as nonpropagating fields that 
are determined by the their equations of motion as functions of the other
fields. A dynamical condensate has been studied~\cite{yy} for the case 
$\G_{hid}=E_8$, and it was shown that the mass of the condensate superfield is
larger than the condensate scale $\Lambda_c$. When this field is
integrated out, the static $E_8$ model of~\cite{us} is recovered.  In
particular, the massless axion is essentially the universal one, up to
$O(m^2_{\tG}/\Lambda^2_c)$ mixing effects.}
is massless~\cite{bd}, and the masses of the dilaton and the complex moduli 
are related to the gravitino mass by
\be m_d \sim {1 \over b_a^2} m_{\tilde G}, \quad
m_{t_I} \approx {2\pi\over3}{(b_{E_8}-b_a)\over(1+b_{E_8}<\ell>)}m_{\tilde G}.
\label{modmass} \ee
In order to generate a hierarchy of order $m_{\tilde G}\sim
10^{-15}m_{Pl}\sim 10^3GeV$ we require~\cite{us} $b_{E_8}/b_a\approx 10$ in 
which case $m_{t_I}\approx 20 m_{\tilde G},\;m_d\sim 
10^3m_{\tilde G}$, which may be sufficient~\cite{us,bcc} to solve the so-called
cosmological moduli problem~\cite{modprob,bd1,lsp}.  Since $m_d\sim
10^6GeV$, its decay does not contribute to the
moduli problem.  The moduli masses are about 20$TeV$, which is sufficient to
evade the late moduli decay problem~\cite{modprob}, but requires R-parity 
violation~\cite{lsp} to avoid a large relic LSP density. If R-parity is 
conserved, this problem can be evaded if the moduli are stabilized at or near 
their vacuum values -- or for a modulus that is itself the inflaton.

In~\cite{lyth}, an explicit model for inflation, based on this effective
theory, was constructed in which the dilaton is stabilized within its domain of
attraction, one or more moduli are stabilized at the vacuum value value 
$t_I=e^{i\pi/6}$, and one of the moduli may be the inflaton.  It is possible 
that the requirement that the remaining moduli be in the domain of attraction 
is sufficient to avoid the problem altogether.   For example, if $\im t_I =
0$, the domain of attraction near $t_I = 1$ is rather limited: 
$0.6<{\rm Re}t_I<1.6$, and the entropy produced by dilaton decay with an 
initial value in this range might be less than commonly assumed. 

If there are several condensates with different $\beta$-functions, the 
potential and the masses (\ref{modmass}) are
dominated by the condensate with the largest $\beta$-function 
coefficient $b_a$, and the result is essentially the same as in the single 
condensate case, except that a small mass is generated for the dynamical
axion.  If there is just one hidden sector condensate,
the axion is massless\footnote{Higher dimension operators might give additional
contributions~\cite{bd} to the axion mass.} up to QCD-induced effects: 
$m_a\sim(\Lambda_{QCD}/\Lambda_c)^{3\over2}m_{\tG}$, and it is the 
natural candidate for the Peccei-Quinn axion. Because of string nonperturbative
corrections to its gauge kinetic term, the decay constant $f_a$ of the 
canonically normalized axion is reduced with respect to the standard result by
a factor $b_a \ell^2 \sqrt{6}\approx 1/50$ if $b_a\approx .1b_{E_8}$, which 
may be sufficiently small to satisfy the (looser) constraints on $f_a$ when 
moduli are present~\cite{bd1}.

\section{Implications for phenomenology and open questions}

The string nonperturbative corrections necessary to stabilize the dilaton
modify the boundary condition (\ref{gstring}) for gauge coupling unification.
Including the functions $f(\ell)$ and $g(\ell)$ we obtain, with $g$ given in 
(\ref{einstein}):
\bea g_a^{-2} (m_s) &=& g^{-2} + {C_a \over 8 \pi^2} \ln (\lambda e)
-{1 \over 16 \pi^2} \sum_I b_a^I \ln (t_I + \bar t_I)|\eta^2(t_I)|^2,
\nonumber \\ m_s^2 &=& \lambda g^2m_{Pl}^2, \quad 
\lambda = {1\over 2}\left< e^{g(\ell)-1} [f(\ell)+1]\right>. \eea
In the perturbative case $\lambda = 1/(2e)\approx .18$,
while a specific fit~\cite{us} with $\alpha = 1/25$ gives a negligible
correction: $\lambda = e^{-1.65} \approx .19$. Another fit~\cite{lyth} with
$\alpha \approx .17$, used to stabilize the dilaton during inflation, gives
$\lambda \approx .15$.

The gaugino masses, as determined at the condensate scale $\Lambda_c$, are
\be m_{\lambda_b}(\Lambda_c) \approx - {3g^2_b(\Lambda_c)b_a\over2\(1+b_a<\ell>
\)}m_{\tG}.\label{gmass}\ee
The one-loop RGE's predict the same ratios among gaugino masses as is 
conventionally assumed, but with absolute values that are below 
experimental bounds if $m_{\tG}\sim $TeV.  However, because the masses
(\ref{gmass}) are negative
(in the phase convention of {\it e.g.}~\cite{twoloop}), they can be driven
to much larger values by two-loop corrections~\cite{twoloop} if there are 
sufficiently massive gauge-charged scalars with large Yukawa couplings.

The soft terms in the scalar potential are sensitive to the -- as yet 
unknown -- details of matter-dependent contributions to string 
threshold corrections and to
the Green-Schwarz term. Neglecting the former,\footnote{If the threshold
corrections are determined by a holomorphic function, they cannot contribute to
scalar masses.} the Green-Schwarz term is
\be V_{GS} = b\sum_Ig^I + \sum_Ap_Ae^{\sum_Iq^A_Ig^I}|\Phi^A|^2 
+ O(|\Phi^A|^4), \quad g^I= -\ln(t^I + \t^I), \ee
and the full K\"ahler potential reads
\be K = \ln(L) + g(L) + \sum_Ig^I  + 
\sum_Ae^{\sum_Iq^A_Ig^I}|\Phi^A|^2 +  O(|\Phi^A|^4).\ee
The cubic ``A-terms'' and scalar masses are given, respectively, by
\bea V_A(\phi) &\approx &e^{K/2}\sum_Am_{\tG}\[\sum_A{p_A-b_a\over1+p_A<\ell>}
\phi^AW_A(\phi) + {3b_a\over1+b_a<\ell>}W(\phi) \] + {\rm h.c.},\nonumber\\  
m_A^2 &\approx& m^2_{\tG}{(p_A - b_a)^2\over(1 + p_A<\ell>)^2}, \eea
where $W(\Phi)$ is the cubic superpotential for chiral
matter. The scalar squared masses are positive and independent of their 
modular 
weights by virtue of the fact that $<F^I>$ vanishes in the vacuum. They
are universal -- and unwanted flavor-changing neutral currents are thereby 
suppressed -- if their couplings to the Green-Schwarz term are universal, in
which case the A-terms reduce to 
\be V_A(\phi) \approx 3m_{\tG}e^{K/2}W(\phi){p_A\(1+2b_a<\ell>\)-b^2_a
<\ell>\over(1+p_A<\ell>)(1+b_a<\ell>)} + {\rm h.c.}. \ee
If the Green-Schwarz term is independent of the matter fields $\Phi^A$,
$p_A = 0$ and we have $m_A = m_{\tilde G},\;A \approx 2m_{\lambda}$.  A
plausible alternative is that the Green-Schwarz term depends only on
the radii $R_I$ of the three compact tori that determine the untwisted sector
part of the K\"ahler potential (17): 
$$ K = \ln(L) + g(L) - \sum_I\ln(2R_I^2) + O(|\Phi^A_{\rm twisted}|^2),$$ where
$2R^2_I = T^I + \bar{T}^I -
\sum_A|\Phi^A_I|^2$ in string units. In this case $p_A = b$ for the
untwisted chiral multiplets $\Phi^A_I$ and the 
untwisted scalars have masses comparable to the moduli 
masses:\footnote{Scenarios in
which the sparticles of the first two generations have masses as high as
$20$ TeV have in fact been proposed~\cite{cohen}.}  
$m_A = m_t/2 \approx A/3$.  If there is a sector with $p_A =
b$ and a Yukawa coupling of order one involving $SU(3)$ triplets ({\it
e.g.} $\bar{D}DN$, where $N$ is a standard model singlet), its two-loop
contribution to gaugino masses can generate gluino masses that are well within
experimental bounds if $m_{\tilde G} \sim$ TeV. Such a coupling could also
generate a $vev$ for $N$, thus breaking possible 
additional $U(1)$'s at a scale $\sim 10$ TeV. The phenomenologically
required $\mu$-term of the MSSM may also be 
generated by the $vev$ of a Standard
Model gauge singlet or by one of the other mechanisms that have been
proposed in the literature~\cite{muterm}.  Finally,
a flat direction af the classical
scalar potential with a mass of $\sim 10 TeV$ would attenuate~\cite{cgmo} 
the baryon dilution problem in the Afflick-Dine mechanism for baryogenesis.

More complete information on
the $\Phi$-dependence of the string scale gauge coupling
functions is required to make precise predictions for soft supersymmetry
breaking.  Nevertheless this model suggests soft supersymmetry breaking
patterns that may differ significantly from those generally assumed in
the context of the MSSM.  Phenomenological constraints such as current
limits on sparticle masses, gauge coupling unification and a charge and
color invariant vacuum~\cite{hit} can be used to restrict the allowed values 
of the $p_A$ as well as the low energy spectrum of the string effective field 
theory. A numerical analysis of these issues is in progress~\cite{grads}.

The soft symmetry breaking parameters given above were calculated for
the CP invariant vacuum $<t_I>=1$.  If some $<t_I>=e^{i\pi/6}$ in the true
vacuum, the results are expected to be similar, except for the possible presence
of CP violating phases.  It remains to be determined whether these effects
can provide the source of the observed CP violation.

In typical orbifold compactifications, the gauge group $\G_{obs}\otimes\G_{hid}
$ obtained at the string scale has no asymptotically free subgroup. However in 
many compactifications with realistic particle spectra~\cite{iban}, the 
effective field theory has an anomalous $U(1)$ gauge subgroup, which is not
anomalous at the string theory level.  The anomaly is cancelled~\cite{dsw} 
by a GS counterterm, similar to the GS term introduced above to cancel the
modular anomaly.  This results in a $D$-term that forces some otherwise
flat direction in scalar field space to acquire a vacuum expectation value,
further breaking the gauge symmetry, and giving large masses to some chiral
multiplets, so that the $\beta$-function of some of the surviving 
gauge subgroups may be negative at lower energy. It has been 
observed~\cite{u1} that $D$-term may play a significant role in 
supersymmetry breaking.  Its presence was explicitly invoked in the 
above-mentioned inflationary model.~\cite{lyth}  Its incorporation into the 
effective condensation potential is under study~\cite{dterm}.

\section{Conclusions}
There have been exciting developments~\cite{duff} in string dualities that
allow the study of a strongly coupled theory by relating it {\it via} a duality
transformation to a different, weakly coupled theory, where perturbative 
methods can be used.  The dilaton runaway problem ($g^2\to 0$) for the weakly 
coupled theory then implies, however, that the strongly coupled theory is dual
to a different weakly coupled theory with the same problem. This has led
to the suggestion~\cite{bd1} that the true vacuum is at strong -- but not
too strong -- coupling, meaning that neither the theory nor its dual is
weakly coupled.  Then instead of perturbation theory, symmetry
arguments, reminiscent of chiral and flavor symmetry arguments 
used to study low energy QCD, must be used to make predictions
for effective field theories from strings. 

The results presented here show that,
to the extent that one can do reliable calculations (in practice in the 
context of orbifold compactification) in effective field theories that satisfy
the known constraints of string theory, weakly coupled string theory is 
compatible with phenomenology, provided string nonperturbative effects are 
taken into account.  These effects can stabilize the dilaton at a weakly 
coupled vacuum value. 

The developments in string dualites have led to the unification of all known 
string theories as different vacua of a single theory: $M$-theory~\cite{duff}.
Among these vacua the weakly coupled heterotic string remains a viable and
attractive possiblity for the description of nature.

\vskip .8cm
\noindent {\bf Acknowledgements}
\vskip .5cm
I wish to acknowledge my collaborators, especially Pierre Bin\'etruy and 
Yi-Yen Wu, and I thank Keith Dienes for discussions.   
This work was supported in part by the Director, Office 
of Energy Research, Office of High Energy and Nuclear Physics, Division of 
High Energy Physics of the U.S. Department of Energy under Contract 
DE-AC03-76SF00098 and in part by the National Science Foundation under 
grant PHY-95-14797.


\begin{thebibliography}{99}
\bibitem{diff} E. Witten, {\it Nucl. Phys.} {\bf B471:} 135 (1996).
\bibitem{bd1} T. Banks and M. Dine, \np {\bf B505:} 445 (1997).
\bibitem{gross} D. Gross, J. Harvey, E. Martinec and R. Rohm, 
{\it Phys. Rev. Lett.} {\bf 54:} 502 (1985).
\bibitem{cy} P. Candelas, G. Horowitz, A. Strominger and E. Witten, 
{\it Nucl. Phys.} {\bf B258:} 46 (1985).
\bibitem{orb} L.J. Dixon, V.S. Kaplunovsky and J. Louis, \np  {\bf B329:} 
27 (1990);
S. Ferrara, D. L\"ust, and S. Theisen, {\it Phys. Lett.} {\bf 233B:} 147 
(1989).
\bibitem{gs} M. Green and J. Schwarz, {\it Phys. Lett.} {\bf 149B:} 117 (1984).
\bibitem{iban} L.E. Iba\~nez, H.-P. Nilles and F. Quevedo, \pl {\bf
187B:} 25 (1987); A. Font, L. Iban\~nez, D. Lust and F.
Quevedo, \pl {\bf 245B:} 401 (1990).  
\bibitem{hos} Y. Hosotani, {\it Phys. Lett.} {\bf 129B:} 75 (1985).
\bibitem{lang} For a review of conventional GUTs, see P. Langacker,
{\it Phys. Rep.} {\bf 72C:} 185 (1981).
\bibitem{cand} P. Candelas, this volume.
\bibitem{sugra} R. Arnowitt, A.H. Chamseddine and P. Nath, \prl {\bf 49:} 970
(1982) and \pl {\bf 121B:} 33 (1983); R. Barbieri, S. Ferrara and C Savoy,
\pl {\bf 199B:} 343 (1982); L. Hall, Lykken and Weinberg, \pr {\bf D27:} 2359 
(1983); see P. Nath, this volume.
\bibitem{nilles}H.P. Nilles, {\it Phys. Lett.} {\bf 115B:} 193 (1982).
\bibitem{wit} E. Witten, {\it Phys. Lett.} {\bf 155B:} 151 (1985).      
\bibitem{dine} M. Dine, R. Rohm, N. Seiberg and E. Witten,
{\it Phys. Lett.} {\bf 156B:} 55 (1985).
\bibitem{kap0} V. Kaplunovski, {\it Phys. Rev. Lett.} {\bf 55} 1036 (1985).
\bibitem{kap2} E. Caceres, V. Kaplunovski and I.M. Mandelberg, \np {\bf 493B:} 
73 (1997). 
\bibitem{duff} M. Duff, this volume.
\bibitem{hw} P. Ho\v rava and E. Witten, \np {\bf B460:}
506 (1996) and {\bf B475:} 94 (1996).
\bibitem{mk} M.K. Gaillard, {\it Phys.Lett.} {\bf 342B:} 125 (1995) and
{\bf 347B:} 284 (1995).
\bibitem{tom} M.K. Gaillard and T.R. Taylor, {\it Nucl. Phys.} {\bf B381:} 577 
(1992).
\bibitem{mod} A. Giveon, N. Malkin and E. Rabinovici, \pl {\bf 220B:} 551
(1989); E. Alvarez and M. Osorio, \pr {\bf D40:} 1150 (1989). 
\bibitem{gsterm} G.L. Cardoso and B.A. Ovrut, \np {\bf B369:} 315 (1993); 
J.-P. Derendinger, S. Ferrara,
C. Kounnas and F. Zwirner, {\it Nucl. Phys.} {\bf B372:} 145 (1992).
\bibitem{thresh} L.J. Dixon, V.S. Kaplunovsky and J. Louis, \np 
{\bf B355:} 649 (1991); I. Antoniadis, K.S. Narain and T.R. Taylor, \pl 
{\bf 267B:} 37 (1991).
\bibitem{kkpr} E. Kiritsis, C. Kounnas, P.M. Petropoulos and
J. Rizos, \np {\bf B483:} 144 (1997).
\bibitem{kl} V.S. Kaplunovsky and J. Louis, \np {\bf B444:} 191 (1995).
\bibitem{linear} S. Ferrara, J. Wess and B. Zumino, {\it Phys. Lett.} {\bf 
51B:} 239 (1974); S. Ferrara and M. Villasante, {\it Phys. Lett.} {\bf 186B:} 85
(1987); S. Cecotti, S. Ferrara and M. Villasante, {\it Int. J. Mod. Phys.} 
{\bf A2} 1839 (1987); S.J. Gates, Jr., P. Majumdar, R. Oerter, and A. E. M. 
van de Ven, {\it Phys. Lett.} {\bf 214B:} 26 (1988); W. Siegel, {\it Phys. 
Lett.} {\bf 211B:} 55 (1988).
\bibitem{bggm} P. Bin\'{e}truy, G. Girardi, R. Grimm and M. M\"{u}ller, 
\pl {\bf 195B:} 83 (1987) and {\bf 265B:} 111 (1991).
\bibitem{deboer} J. de Boer and K. Skenderis, {\it Nucl. Phys.} {\bf B481:} 
129 (1996); S.J. Gates, private communication.
\bibitem{us} P. Bin\'{e}truy, M. K. Gaillard and Y.-Y. Wu, {\it Nucl. Phys.}
{\bf B481:} 109 (1996) and {\bf B493:} 27 (1997); 
{\it Phys. Lett.} {\bf 412B:} 228 (1997).
\bibitem{bdqq} C.P. Burgess, J.-P. Derendinger, F. Quevedo and M. Quir\'{o}s,
{\it Phys. Lett.} {\bf 348B:} 428 (1995).
\bibitem{kap} V.S. Kaplunovski, \np {\bf B307:} 145 (1988); Erratum: 
{\it ibid}., {\bf B382:} 436 (1992).
\bibitem{unif} L.E. Iba\~nez, D. L\"ust and G.G. Ross, {\it Phys. Lett.}
{\bf 272B:} 251 (1991) ;  
L.E. Iba\~nez and D. L\"ust, \np {\bf B382:} 305 (1992);  
M. K. Gaillard, and R. Xiu, {\it Phys. Lett.} {\bf 296B:} 71 (1992);  
R. Xiu, {\it Phys. Rev.} {\bf D49:} 6656 (1994);  
S. P. Martin and P. Ramond, {\it Phys. Rev.} {\bf D51:} 6515 (1995). 
K.R. Dienes and A.E. Farraggi, {\it Phys. Rev. Lett.} {\bf 75} 2646 (1995);
see K.R. Dienes, {\it Phys. Rep.} {\bf 287C:} 447 (1997)
for a review and further references.
\bibitem{aff} K.R. Dienes, A.E. Faraggi, and J. March-Russell,
\np {\bf B467} 44 (1996); S. Chaudhuri, S.-W.Chung, G. Hockney and J. Lykken, 
\np {\bf B469}, 357 (1996). 
\bibitem{rw} R. Rohm and E. Witten {\it Ann. Phys.} {\bf 170:} 454 (1986).
\bibitem{bg} P. Bin\'{e}truy and M.K. Gaillard, 
{\it Phys. Lett.} {\bf 232B:} 82 (1989). 
\bibitem{shenk} S.H. Shenker, in {\it Random Surfaces
and Quantum Gravity}, Proceedings of the NATO Advanced Study
Institute, Cargese, France, 1990, edited by O. Alvarez,
E. Marinari, and P. Windey, NATO ASI Series B: Physics Vol.262
(Plenum, New York, 1990).
\bibitem{casas} J.A. Casas, \pl {\bf 384B:} 103 (1996).
\bibitem{bgt} P. Bin\'{e}truy, M.K. Gaillard and
T.R. Taylor, {\it Nucl. Phys.} {\bf B455:} 97 (1995).
\bibitem{vy} G. Veneziano and S. Yankielowicz,
\pl {\bf 113B:} 231 (1982); T.R. Taylor, G. Veneziano and S. Yankielowicz,
\np {\bf B218:} 493 (1983).
\bibitem{bg96} 
P. Bin\'etruy and M.K. Gaillard, {\it Phys.\ Lett.} {\bf 365B:} 87 (1996).
\bibitem{race} N.V. Krasnikov, {\it Phys. Lett.} {\bf 193B:} 37 (1987); 
J.A. Casas, Z. Lalak, C. Mu\~noz, and G.G. Ross, {\it Phys. Lett.} {\bf 347B:} 
243 (1990); B. de Carlos, J.A. Casas and C. Mu\~noz,  {\it Phys. Lett.} 
{\bf 399B:} 623 (1993). 
\bibitem{lyth}M. K. Gaillard, H. Murayama and D.H. Lyth, preprint
LANCS-TH/9705, LBNL-40830, UCB-PTH-97/48 (1998).
\bibitem{eva} E. Silverstein, \pl {\bf 396B:} 91 (1997).
\bibitem{bd} T. Banks and M. Dine, {\it Phys. Rev.} {\bf D50:} 7454 (1994).
\bibitem{yy} Y.-Y. Wu, Ph. D  Thesis, hep-th/9610089 and
Berkeley preprint, hep-th/9610089.
\bibitem{bcc} T. Barreiro, B. de Carlos, E.J. Copeland, SUSX-TH-97-024, 
hep-ph/9712443 (1997).
\bibitem{modprob} G.D. Coughlan, W. Fischler, E.W. Kolb, S. Raby and G.G.
 Ross, \pl {\bf 131B:} (1983) 59;
\bibitem{lsp} M. Kawasaki, T. Moroi, and T. Yanagida, \pl
{\bf 370B:}, 52 (1996).
\bibitem{twoloop} S.P. Martin and M.T. Vaughn, \pl {\bf 318B:} 331
(1993), and \pl {\bf D50:} 2282 (1994);   
Y. Yamada, {\it Phys. Rev. Lett.} {\bf 72} 25 (1994).
\bibitem{cohen} A.G. Cohen, D.B. Kaplan and A.E. Nelson, \pl {\bf 388B:}
588 (1996);  A. Pomarol and D. Tommasini, \np {\bf 466B:} 3 (1996);  
G. Dvali and A. Pomarol, {\it Phys. Rev. Lett.} {\bf 77:} 3728 (1996).
\bibitem{muterm} G.F. Giudice, and A. Masiero, \pl {\bf 206B:} 480 (1988);   
J. A. Casas and C. Mun\~oz, \pl {\bf 306B:} 288 (1993);
I. Antoniadis, E. Gava, K.S. Narain and T.R. Taylor, \pl {\bf B432} 
187 (1994).
\bibitem{cgmo} B.A. Campbell, M.K. Gaillard, H. Murayama and K.A. Olive, 
preprint LBNL-41768, UCB-PTH-98/23, hep-ph/9805300 (1998).
\bibitem{hit} N. Arkani-Hamed and H. Murayama, \pr {\bf D56:} 6733 (1997);
K. Agashe and M. Graesser, Berkeley preprint hep-ph/9801446.
\bibitem{grads} B. Nelson and D. Smith, in progress.
\bibitem{dsw}  M. Dine, N. Seiberg and E. Witten, \np {\bf B289:}
589 (1987); J. Attick, L. Dixon and A. Sen, \np {\bf B292:} 109 (1987);
M. Dine, I. Ichinose and N. Seiberg, \np {\bf B293:} 253 (1988).
\bibitem{u1} P. Bin\'etruy and E. Dudas, \pl {\bf 389B:} 503 (1996).
\bibitem{dterm} P. Bin\'etruy and M.K. Gaillard, in progress.
\end{thebibliography}
\end{document}